\documentclass[
  aps,
  prl,
  twocolumn,
  superscriptaddress,
  nofootinbib
]{revtex4-2}

\usepackage{amsmath,amssymb,amsfonts}
\usepackage{graphicx}
\usepackage{xcolor}
\usepackage{hyperref}

\hypersetup{
  colorlinks=true,
  linkcolor=blue,
  citecolor=blue,
  urlcolor=blue
}

\begin{document}

\title{Uniqueness of Static Black Holes in Lovelock Gravity}

\author{Chanchal Sharma}
\affiliation{
Department of Physics,
Indian Institute of Technology Gandhinagar,
Palaj, Gandhinagar 382055, India
}
\email{chanchal.sharma@iitgn.ac.in \\ sudiptas@iitgn.ac.in}

\author{Sudipta Sarkar}
\affiliation{
Department of Physics,
Indian Institute of Technology Gandhinagar,
Palaj, Gandhinagar 382055, India
}

\date{\today}

\begin{abstract}
Black hole uniqueness is one of the fundamental predictions of general relativity: the spherically symmetric Schwarzschild--Tangherlini spacetime is the unique regular, asymptotically flat, static vacuum black hole solution of Einstein's equations. We show that an analogous result holds perturbatively in Lovelock gravity. Assuming the appropriate positive-mass rigidity statement, any regular, asymptotically flat, static vacuum black hole solution of Lovelock gravity on a generic branch admitting a regular Taylor expansion about general relativity is uniquely determined, at fixed ADM mass, by the corresponding spherically symmetric Lovelock black hole solution. In particular, Einstein--Gauss--Bonnet gravity admits no distinct nonspherical static branch perturbatively connected to Schwarzschild--Tangherlini, and the unique solution is Boulware--Deser. The same recursive mechanism extends to the full Lovelock class. Our result shows that static black hole uniqueness can extend beyond general relativity and is not a property peculiar to the Einstein equations.
\end{abstract}

\maketitle

{\bf \textit{Introduction.}}
Black hole uniqueness is one of the most remarkable structural results of general relativity. For a regular, asymptotically flat vacuum black hole, staticity is an extremely restrictive condition: it forces spherical symmetry, and the geometry is uniquely Schwarzschild--Tangherlini \cite{Israel1967,BuntingMasood1987,Tangherlini1963,GibbonsIdaShiromizu2002}. This is much stronger than merely knowing an exact spherical solution. Uniqueness is what turns the Schwarzschild geometry into a prediction of the theory rather than simply one member of a larger family of possible black holes.

A natural question is whether this remarkable rigidity survives beyond Einstein gravity. There is no obvious reason that it should. Higher-curvature interactions modify the nonlinear field equations, can introduce additional branches of solutions, and may alter basic properties of black hole spacetimes. Lovelock gravity provides a particularly clean setting in which to ask this question. It is the most general metric theory with second-order field equations despite containing higher powers of curvature \cite{Lovelock1971}. Its first nontrivial correction is the Gauss--Bonnet interaction, while higher Lovelock densities generate increasingly nonlinear curvature terms.

Spherically symmetric black holes in these theories have been known for decades. In Einstein--Gauss--Bonnet gravity the corresponding solution is the Boulware--Deser black hole \cite{BoulwareDeser1985}, and more general spherical Lovelock solutions are determined by an algebraic Lovelock polynomial \cite{Wheeler1986,Zegers2005}. But the existence of these solutions does not establish uniqueness. In particular, it does not exclude the possibility of additional regular static black holes with nonspherical geometry but the same asymptotic charges. The central question is therefore simple: \emph{does staticity continue to imply spherical symmetry once genuine higher-curvature interactions are present?}

In this Letter we show that, on the regular branch continuously connected to general relativity, the answer is yes. Subject to the appropriate positive-mass rigidity statement, every regular, asymptotically flat, static vacuum black hole in Einstein--Gauss--Bonnet gravity is perturbatively forced to be spherically symmetric to all orders in the coupling. At fixed ADM mass, the unique solution is therefore the Einstein branch of the Boulware--Deser black hole. We then show that the same recursive mechanism extends to general Lovelock gravity. We present the Einstein--Gauss--Bonnet case explicitly in the main text, while the extension to general Lovelock gravity is given in the End Matter. Thus no distinct static nonspherical black hole branch can bifurcate perturbatively from Schwarzschild--Tangherlini.

The proof is built around the same deep geometric ingredient that underlies static uniqueness in GR: positive-mass rigidity. In the Einstein theory, a conformal transformation and doubling across the horizon produce a complete asymptotically flat spatial geometry of zero ADM mass; the equality case of the positive-mass theorem then forces this geometry to be flat \cite{BuntingMasood1987,GibbonsIdaShiromizu2002}. Establishing an equally general positive-mass theorem in Lovelock gravity is itself a difficult open problem. Our result isolates precisely this global input: assuming the corresponding zero-mass rigidity statement, we show that the rest of the uniqueness argument survives the higher-curvature corrections.

The mechanism is perturbative but extends to all orders. We restrict to the Einstein-connected branch and assume a regular Taylor expansion in the Lovelock couplings. A branch-adapted conformal transformation gives a doubled geometry whose ADM mass vanishes exactly. If the first possible deformation occurs at order $\alpha^N$, the higher-curvature field equations imply that the generalized curvature of the conformal geometry is $O(\alpha^{N+1})$. Hence, in the theory consistently truncated at $O(\alpha^N)$, the transformed geometry lies in the zero-mass equality sector. Positive-mass rigidity then excludes any non-gauge deformation at that order. Repeating the argument fixes the conformal geometry recursively, while the remaining static equations force spherical symmetry.

This result adds black hole uniqueness to a growing class of strong geometric properties of general relativity that survive on perturbative Einstein-connected branches of higher-curvature theories. Related examples include the black hole zeroth law and circularity in broad classes of higher-derivative gravitational theories \cite{GhoshSarkarZeroth,BhattacharyyaZeroth,XieCircularity}. Uniqueness is, however, a more global statement: it constrains the complete exterior geometry rather than only the horizon or the integrability properties of the spacetime.

Our theorem is local in coupling space. It applies to generic Lovelock couplings on the regular Taylor-expandable Einstein-connected branch and does not constrain disconnected, nonanalytic, singular, or degenerate branches. Nevertheless, it shows that the fundamental GR statement that staticity implies spherical symmetry is not peculiar to the Einstein equations, but persists throughout a nontrivial class of genuine higher-curvature gravitational theories.

{\bf \textit{Static black hole uniqueness in GR.}}
We briefly recall the conformal proof in GR
\cite{BuntingMasood1987,GibbonsIdaShiromizu2002}. For a general static metric
\begin{equation}
 ds^2=-V^2dt^2+h_{ij}dx^idx^j ,
 \label{eq:static}
\end{equation}
the vacuum equations give
\begin{equation}
 D^2V=0,
 \qquad
 R_{ij}=V^{-1}D_iD_jV,
 \label{eq:static_GR_eqs}
\end{equation}
and hence $R[h]=0$. Introducing
\begin{equation}
 \widetilde h^{\,\pm}_{ij}
 =
 U_\pm^{4/(D-3)}h_{ij},
 \qquad
 U_\pm=\frac{1\pm V}{2},
 \label{eq:GR_conformal}
\end{equation}
one finds
\begin{equation}
\begin{aligned}
 \widetilde R_\pm=U_\pm^{-\frac{D+1}{D-3}}\bigg[&U_\pm R[h]-\frac{4(D-2)}{D-3}D^2U_\pm\bigg],
\end{aligned}
\label{eq:scalar_transform}
\end{equation}
so $\widetilde R_\pm=0$. The two copies join smoothly across the horizon, the minus copy compactifies its asymptotic end, and the ADM mass of the complete doubled manifold vanishes. Positive-mass rigidity therefore gives $\widetilde h_{ij}=\delta_{ij}$. The remaining static equations and boundary conditions then force the level surfaces of $V$ to be round and concentric, yielding Schwarzschild--Tangherlini. Thus the GR proof rests on two ingredients: a conformal doubling with zero ADM mass and rigidity of the corresponding scalar-curvature equality case.

{\bf \textit{Einstein--Gauss--Bonnet gravity.}}
The EGB action is
\begin{equation}
 I=
 \frac{1}{16\pi G_D}
 \int d^Dx\,\sqrt{-g}
 \left(R+\alpha{\cal L}_{\rm GB}\right),
 \label{eq:EGBaction}
\end{equation}
with
\begin{equation}
 {\cal L}_{\rm GB}
 =
 R_{abcd}R^{abcd}
 -4R_{ab}R^{ab}
 +R^2 .
\end{equation}
The vacuum equations are $G_{ab}+\alpha H_{ab}=0$. Their spherically symmetric solution is the Boulware--Deser geometry
\cite{BoulwareDeser1985},
\begin{equation}
    ds^2=-f_\pm dt^2+\frac{dr^2}{f_\pm}+r^2d\Omega_{D-2}^2 ,
\end{equation}
where
\begin{equation}
 f_\pm
 =
 1+\frac{r^2}{2\hat\alpha}
 \left[
 1\pm
 \sqrt{1+\frac{4\hat\alpha\mu}{r^{D-1}}}
 \right],
 \label{eq:BDmetric}
\end{equation}
with $\hat\alpha=(D-3)(D-4)\alpha$ and $\mu$ as the black hole mass. We restrict to the minus branch, which has a smooth Einstein limit.

The vacuum time-symmetric Hamiltonian constraint for a general static metric is
\begin{equation}
{\cal Q}[h]\equiv R[h]+\alpha{\cal L}_{\rm GB}[h]=0 .
 \label{eq:EGBHamiltonian}
\end{equation}
We assume the corresponding exact positive-mass statement for the relevant class of asymptotically flat spatial geometries, in particular its equality-case rigidity: ${\cal Q}\geq0$ implies non-negative ADM mass, with zero ADM mass implying flatness modulo diffeomorphisms. This statement is not known for general EGB initial data; existing results require additional assumptions \cite{ShiromizuOhashi2013,GeWangWu2014}. Our result is therefore conditional on this assumption.

We additionally restrict to the regular Taylor-expandable Einstein-connected branch. This assumption is essential for the perturbative use of rigidity because it identifies a definite first order at which a departure from the reference geometry can occur. The conformal construction used below has $\widetilde M_{\rm ADM}=0$ exactly. If the first possible deformation occurs at order $\alpha^N$ and the field equations give
\begin{equation}
 \widetilde{\cal Q}=O(\alpha^{N+1}),
\end{equation}
then in the theory consistently truncated at $O(\alpha^N)$ one has $\widetilde{\cal Q}=0$ while the ADM mass remains exactly zero. The zero-mass equality-case rigidity therefore excludes any physical nonflat deformation at that order. Higher-order terms cannot alter a lower-order Taylor coefficient. This is the order-by-order use of the exact rigidity statement adopted below. Known dynamical instabilities of some Boulware--Deser black holes do not by themselves contradict positive-mass rigidity \cite{DottiGleiser2005,BeroizDottiGleiser2007}.

{\bf \textit{Perturbative uniqueness in EGB gravity.}}
We now implement the conformal construction for the general static geometry given by Eq.~(\ref{eq:static}). The proof proceeds in three steps: construct a zero mass conformal doubling, show that its generalized curvature vanishes order by order, and then use the remaining static equations to recover spherical symmetry.

The conformal factor is suggested by writing the Einstein branch of the Boulware--Deser geometry in isotropic coordinates, which gives lapse-dependent factor as
\begin{equation}
\begin{aligned}
 \Omega_+(V,\alpha;\mu)
 ={}&\frac{1+V}{2}
 \exp\Bigg\{
 \int_1^V\frac{du}{1+u}
 \\
 &\times
 \frac{(D-1)\widehat\alpha(1-u^2)}
 {(D-3)r_{\rm BD}^2(u)
 +(D-5)\widehat\alpha(1-u^2)}
 \Bigg\},
\end{aligned}
\label{eq:OmegaEGB}
\end{equation}
where $r_{\rm BD}(V)$ is implicitly defined by
\begin{equation}
    \begin{aligned}
        \mu={}&(1-V^2)r_{\rm BD}^{D-3}+\widehat\alpha(1-V^2)^2 r_{\rm BD}^{D-5}.
    \end{aligned}
\label{eq:rBDV}
\end{equation}
Other conformal factor is $\Omega_-(V,\alpha;\mu)=\Omega_+(-V,\alpha;\mu)$. As $\alpha\rightarrow 0$, we get $\Omega_+=U_+$.

The Boulware--Deser geometry is used only to determine the conformal factor. We then apply it to the completely general static metric,
\begin{equation}
 \widetilde h^{\,\pm}_{ij}
 =
 \Omega_\pm^{4/(D-3)}h_{ij},
 \label{eq:EGBconformal}
\end{equation}
without imposing any symmetry on $V$ or $h_{ij}$.

The two conformal factors agree at the horizon, while the minus copy compactifies the second asymptotic end. Asymptotically, $1-V\simeq\mu/(2r^{D-3})$, whereas the explicit Gauss--Bonnet corrections to $\Omega_+$ fall faster than $r^{-(D-3)}$. They therefore do not affect the ADM mass coefficient. The coefficient of the $r^{-(D-3)}$ term cancels exactly as in the GR Bunting construction. Thus, at fixed physical mass, the doubled conformal geometry exactly has
\begin{equation}
 \widetilde M_{\rm ADM}=0.
 \label{eq:Mtildezero}
\end{equation}
For the conformally related metric, we define $\widetilde{\cal Q}\equiv{\cal Q}[\widetilde{h}]$. 

At zeroth order, the GR uniqueness theorem fixes the background to Schwarzschild--Tangherlini, and the transformed metric is Euclidean. Expanding the conformal factor in $\alpha$, we obtain
\begin{equation}
 \Omega(V(\alpha),\alpha)=\frac{1+V(\alpha)}{2}+\alpha\,\omega_1(V_0)+O(\alpha^2),
\end{equation}
with $\Omega_0=U_+(V_0)$. The scalar curvature transforms according to the identity in Eq.~(\ref{eq:scalar_transform}) evaluated with the EGB conformal factor $\Omega_\pm(V,\alpha)$. The first-order piece that enters $\widetilde R$ is
\begin{equation}
\begin{split}
  \widetilde{R}^{(1)} = \Omega_0^{-\frac{D+1}{D-3}} \biggl[ &\Omega_0 R^{(1)}[h] \\
  &- c_D \left( \tfrac{1}{2}(D^2V)^{(1)} + (D_0^2\omega_1)^{(0)} \right) \biggr].
\end{split}
\label{eq:Rtilde1}
\end{equation}
Here, $(D^2V)^{(1)}$ includes both the variation of $V$ and that of the spatial connection. The zeroth order contribution to $\widetilde R$ vanishes because of the Bunting scalar-flatness condition. The Hamiltonian and the trace constraints of the EGB field equations fix $R^{(1)}[h]$ and $(D^2V)^{(1)}$ entirely in terms of the Schwarzschild-Tangherlini background, independent of any unconstrained first-order metric correction. Since $\omega_1$ is a function only of
$V_0$,
\begin{equation}
\begin{aligned}
 (D_0^2\omega_1)^{(0)}
 &=
 \omega_1'(V_0)D_0^2V_0
 +\omega_1''(V_0)(D_0V_0)^2
 \\
 &=
 \omega_1''(V_0)(D_0V_0)^2 .
\end{aligned}
\label{eq:omegaLaplace}
\end{equation}
Thus, every term in ${\widetilde R}^{(1)}$ is determined entirely by the Schwarzschild--Tangherlini background and the branch-adapted function $\omega_1(V_0)$. In particular, ${\widetilde R}^{(1)}$ is the same universal background functional for every regular static candidate sharing this zeroth-order geometry. The isotropic conformal factors maps the spatial geometry of Boulware--Deser representative exactly to Euclidean space, so
\begin{equation}
 {\widetilde R}^{(1)}_{\rm BD}=0,
\end{equation}
Hence follows,
\begin{equation}
 \widetilde R^{(1)}=0.
 \label{eq:Rtilde1zero}
\end{equation}
for an arbitrary static candidate as the zeroth-order conformal prefactor in Eq.~(\ref{eq:Rtilde1}) is non-zero. Since the zeroth-order conformal geometry is flat,
$\widetilde{\cal L}_{\rm GB}^{(0)}=0$, and hence $ \widetilde{\cal Q}^{(1)}=0.$
The regular Taylor expansion gives
\begin{equation}
 \widetilde h_{ij}
 =
 \delta_{ij}
 +\alpha \, k^{(1)}_{ij}
 +O(\alpha^2).
 \label{eq:tildeTaylorFirst}
\end{equation}
These equations imply $\widetilde{\cal Q}=O(\alpha^2)$.  $\widetilde{\cal Q}=0$. At the same time, we have, $\widetilde M_{\rm ADM}=0$ exactly. Therefore, in the theory consistently truncated at $O(\alpha)$, any non-gauge $k^{(1)}_{ij}$ would be a genuine non-flat, zero-mass configuration of the $O(\alpha)$ theory, in direct conflict with the assumed equality-case rigidity. Higher-order terms cannot remove this first-order Taylor coefficient. Hence, if the positive mass theorem holds for EGB gravity, we must have $ k^{(1)}_{ij}
 = {\cal L}_{\xi_1}\delta_{ij}$, and after fixing the asymptotic gauge,
\begin{equation}
 \widetilde h_{ij}
 =
 \delta_{ij}+O(\alpha^2).
 \label{eq:flatOalpha}
\end{equation}

The second-order step is straightforward. Equation~(\ref{eq:flatOalpha}) gives $\widetilde R_{ijkl}=O(\alpha^2)$, so $\alpha\widetilde{\cal L}_{\rm GB}=O(\alpha^5)$ and hence $\widetilde{\cal Q}^{(2)}=\widetilde R^{(2)}$. Since all lower-order coefficients are already fixed, the first $O(\alpha^2)$ difference between an arbitrary static solution and the Boulware--Deser solution satisfies the homogeneous linearized Einstein equation on the Schwarzschild--Tangherlini background. At the same order, the conformal transformation reduces to the linearized GR Bunting map. Therefore the two solutions have the same $\widetilde R^{(2)}$. The Boulware--Deser representative is mapped exactly to flat space, so $\widetilde R^{(2)}=\widetilde{\cal Q}^{(2)}=0$. Thus $\widetilde{\cal Q}=O(\alpha^3)$. In the theory truncated at $O(\alpha^2)$, we therefore have $\widetilde{\cal Q}=0$ and $\widetilde M_{\rm ADM}=0$. Equality-case rigidity then excludes any physical $O(\alpha^2)$ deformation, so after gauge fixing
\[
\widetilde h_{ij}=\delta_{ij}+O(\alpha^3).
\]

The same mechanism iterates. Suppose
\begin{equation}
 \widetilde h_{ij}
 =
 \delta_{ij}
 +\alpha^N k^{(N)}_{ij}
 +O(\alpha^{N+1})
 \label{eq:firstunfixed}
\end{equation}
at the first potentially non-vanishing order. By the induction hypothesis all lower-order coefficients have already been fixed to the Boulware--Deser reference solution. The difference between two candidate solutions at this first unfixed order therefore satisfies the homogeneous linearized Einstein equation on Schwarzschild--Tangherlini and every Gauss--Bonnet contribution carries an additional explicit power of $\alpha$. At the same order the difference of the full conformal maps reduces to the linearized GR Bunting map, so  $\widetilde R^{(N)}=0.$ Moreover, $\widetilde{\rm Riem}=O(\alpha^N)$, and therefore the explicit Gauss--Bonnet term contributes at $\mathcal{O}(\alpha^{N+1})$ or higher. Since all lower coefficients of $\widetilde{\cal Q}$ vanish by the induction hypothesis, we obtain
\begin{equation}
 \widetilde{\cal Q}=O(\alpha^{N+1}).
 \label{eq:QinductiveOrder}
\end{equation}
The ADM mass, by contrast, vanishes exactly:
$\widetilde M_{\rm ADM}=0$. The regular Taylor expansion is crucial here. Since $N$ is by construction the first order at which the conformal geometry could depart from the flat equality solution, the theory consistently truncated
at $O(\alpha^N)$ has $ \widetilde{\cal Q}=0, \, \widetilde M_{\rm ADM}=0.$ A non-gauge $k^{(N)}_{ij}$ would therefore be a non-flat zero-mass equality configuration of the truncated theory, in direct conflict with the assumed rigidity. Terms of order $\alpha^{N+1}$ and higher are not part of this truncation and, in a regular Taylor expansion, cannot remove a nonzero coefficient already present at order $\alpha^N$. Hence, we must have,
\begin{equation}
 \widetilde h_{ij}
 =
 \delta_{ij}+O(\alpha^{N+1}).
\end{equation}
Since $N$ is arbitrary,
\begin{equation}
 \widetilde h_{ij}=\delta_{ij}
 \label{eq:flatallorders}
\end{equation}
to all orders in the regular Taylor expansion.

It remains to show that flatness of the conformal metric fixes the lapse. Let $\rho$ be proper distance from the bifurcation surface on the physical static slice. Locally, $h=d\rho^2+q_{AB}(\rho,x)dx^Adx^B$ and $V=\kappa(x)\rho+O(\rho^3)$. Smooth reflection through the bifurcation surface makes $q_{AB}$ even in $\rho$, so $K_{AB}=\frac12\partial_\rho q_{AB}=0$ on the horizon.

Under the conformal transformation,
\begin{equation}
\begin{aligned}
 \widetilde K_{AB}=\Omega^{2/(D-3)}\bigg[K_{AB}+\frac{2}{D-3}q_{AB}\,n(\ln\Omega)\bigg].
\end{aligned}
\label{eq:Ktransform}
\end{equation}
Thus $\widetilde K_{AB}\propto\widetilde q_{AB}$ on the horizon. Since $\widetilde h_{ij}$ is Euclidean, the horizon is a compact totally umbilical hypersurface and hence a round $S^{D-2}$.

The remaining static equations propagate this roundness away from the horizon. At the first possible nonspherical perturbative order the difference from the spherical reference solution obeys the homogeneous linearized Einstein equation. As shown explicitly in the End Matter, the tangential traceless and mixed projections reduce to
\begin{equation}
 \frac{V-1}{V+1}H_{AB}^{\rm TF}=0,
 \qquad
 \frac{V-1}{V+1}H_{nA}=0,
 \label{eq:Hconditions}
\end{equation}
where $H_{ij}=\partial_i\partial_jV$ is the Hessian in the flat conformal geometry. Since $0\leq V<1$ at every finite point of the exterior, the coefficient is nonzero there. Hence
\begin{equation}
 H_{AB}^{\rm TF}=0,
 \qquad
 H_{nA}=0.
\end{equation}
The first relation removes the traceless deformation of each connected compact equipotential surface at the first nonspherical order; the second removes the angular variation of its normal displacement. Starting from the round horizon, the first possible nonspherical coefficient therefore vanishes. Repeating the argument order by order leaves only concentric round equipotential surfaces, so $V=V(r)$ to all orders in the Taylor expansion.

We have therefore shown that positive-mass rigidity leaves no room for a nonspherical static deformation within the regular Taylor-expandable Einstein-connected branch. The spacetime is spherically symmetric, and the vacuum EGB equations then select the asymptotically flat Einstein-connected Boulware--Deser branch. Thus, at fixed ADM mass, any regular static black hole admitting a Taylor expansion about Schwarzschild--Tangherlini agrees with Boulware--Deser order by order in the Gauss--Bonnet coupling. The same calculation strategy extends to general Lovelock gravity, with details given in the End Matter.

{\bf \textit{Discussion.}}
We have shown that, on the regular Einstein-connected branch, staticity continues to imply spherical symmetry in Lovelock gravity, provided the appropriate positive-mass rigidity theorem holds. Although the spherically symmetric Lovelock solutions are well known, their existence does not exclude distinct nonspherical static geometries. Our result establishes that, within the regular Taylor-expandable sector, spherical symmetry follows from the field equations rather than being imposed as an ansatz.

An important qualification is the positive-mass rigidity assumption, which is not yet established in the required generality even for Einstein--Gauss--Bonnet gravity. Our argument requires its zero-mass equality case. The conformal doubling has exactly vanishing ADM mass, while the field equations ensure that the generalized curvature vanishes at the first unfixed perturbative order. The assumed rigidity then rules out any nontrivial physical deformation at that order. Repeating this step recursively fixes the geometry, order by order, to the corresponding spherically symmetric Lovelock black hole. In this sense, our result isolates zero-mass rigidity as the precise global ingredient required for static black hole uniqueness to survive beyond general relativity.

Our result complements other rigidity properties that survive higher-curvature corrections. The black hole zeroth law holds perturbatively in Lovelock gravity and more general higher-derivative theories \cite{GhoshSarkarZeroth,BhattacharyyaZeroth}, while circularity persists perturbatively in a broad class of gravitational effective theories \cite{XieCircularity}. Uniqueness is a stronger global statement, fixing the entire exterior geometry rather than only the horizon data or the integrability structure of the spacetime.

The theorem is local in coupling space and does not constrain disconnected, nonanalytic, or degenerate Lovelock branches. Whether a non-perturbative uniqueness theorem can be established remains open, particularly in view of recent non-perturbative results concerning the zeroth law in higher-curvature gravity \cite{GhoshSarkarUniversal}. An extension to more general higher-derivative theories would also be valuable.

Rotation presents another natural direction. Unlike the static problem, higher-dimensional GR admits distinct rotating black objects with the same asymptotic charges. Perturbative rotating EGB solutions and numerical black hole and black-ring configurations are already known \cite{KonoplyaZhidenko2020,KleihausKunzRadu2023}. The relevant question is therefore whether an Einstein-connected rotating solution is perturbatively unique within a specified horizon topology and branch. Such a result could constrain the possible departures from the standard black hole geometries relevant for multipole moments, tidal response, and gravitational-wave ringdown.

{\bf \textit{Acknowledgment.}}
We thank Amitabh Virmani, Daisuke Ida, Nitesh Dubey, Sanved Kolekar, and Sumanta Chakraborty for valuable discussions and helpful input. S.S.’s research is supported by the Department of Science and Technology, Government of India, under the ANRF CRG Grant (No. CRG/2023/000934).

\section{End Matter: Extension to general Lovelock gravity.}

We give here further details of the perturbative construction and show how the mechanism developed explicitly for EGB gravity extends to a generic Lovelock theory. Consider
\begin{equation}
 I=\frac{1}{16\pi G_D}
 \int d^Dx\,\sqrt{-g}
 \sum_{p=0}^{K}\alpha_p{\cal L}_p ,
 \qquad
 K=\left\lfloor\frac{D-1}{2}\right\rfloor ,
 \label{eq:LovelockAction}
\end{equation}
with ${\cal L}_1=R$, $\alpha_1=1$, and $\alpha_0=0$ for asymptotic flatness. We restrict to generic Lovelock couplings and to the Einstein-connected branch. For bookkeeping, write
\begin{equation}
 \alpha_p=\epsilon\beta_p,
 \qquad p\ge2,
 \label{eq:LovelockBookkeeping}
\end{equation}
with the $\beta_p$ fixed, and assume that the branch admits a regular Taylor expansion in $\epsilon$ about the Einstein solution. Since the $\beta_p$ are arbitrary, proving the result for arbitrary directions in coupling space determines, order by order, the corresponding multivariate Taylor expansion in the independent Lovelock couplings.

The static spherically symmetric Einstein-connected solution is characterized by the Lovelock polynomial
\begin{equation}
 \mu=
 \sum_{p=1}^{K}
 \widehat\alpha_p
 (1-V^2)^p
 r_L^{D-1-2p},
 \label{eq:LovelockPolynomial}
\end{equation}
where $\widehat\alpha_p$ denotes the conventional dimension-dependent normalization of the Lovelock coupling \cite{Wheeler1986,Zegers2005}. Writing this reference solution in isotropic coordinates determines a lapse-dependent conformal factor. Defining $z=1-V^2$ and
\begin{equation}
 A_p=
 \widehat\alpha_p z^p r_L^{D-1-2p},
\end{equation}
one obtains
\begin{equation}
 \frac{d\ln\Omega_+}{dV}
 =
 \frac{D-3}{1+V}
 \frac{\sum_p p\,A_p}
 {\sum_p(D-1-2p)A_p}.
 \label{eq:LovelockOmegaDerivative}
\end{equation}
Separating the Einstein Bunting factor gives
\begin{equation}
\begin{aligned}
 \frac{d}{dV}
 \ln\!\left(\frac{2\Omega_+}{1+V}\right)
 =
 \frac{1}{1+V}
 \frac{
 (D-1)\sum_{p\ge2}(p-1)A_p
 }{
 \sum_p(D-1-2p)A_p
 } .
\end{aligned}
\label{eq:LovelockOmegaEinstein}
\end{equation}
The factor $(p-1)$ makes the Einstein contribution drop out identically. For $p=2$, Eq.~(\ref{eq:LovelockOmegaEinstein}) reduces to the EGB conformal factor used in the main text.

The higher-Lovelock corrections to $\Omega_+$ fall faster than the Einstein mass term at spatial infinity and therefore do not alter the ADM coefficient. The leading mass term cancels exactly as in the Bunting construction. Thus the doubled conformal geometry satisfies
\begin{equation}
 \widetilde M_{\rm ADM}=0
 \label{eq:LovelockZeroMass}
\end{equation}
exactly.

The perturbative argument now proceeds exactly as in the EGB case. Suppose that the conformal geometry has already been fixed through order
$\epsilon^{N-1}$,
\begin{equation}
 \widetilde h_{ij}
 =
 \delta_{ij}+O(\epsilon^N).
 \label{eq:LovelockInductionAssumption}
\end{equation}
By the induction hypothesis, all lower-order coefficients have already been fixed to those of the spherical Einstein-connected reference solution. At the first unfixed order their difference satisfies the homogeneous linearized Einstein equation on the Schwarzschild--Tangherlini background. Indeed, every Lovelock term with $p\ge2$ carries at least one explicit power of $\epsilon$, so the variation produced by an $O(\epsilon^N)$ difference can first enter at $O(\epsilon^{N+1})$.

At the same order, all explicit Lovelock corrections to the conformal factor are common to the two solutions. Their difference therefore reduces to the linearized GR Bunting transformation. Consequently,
\begin{equation}
 \Delta\widetilde R^{(N)}=0.
\end{equation}
By construction of $\Omega_+$ from the isotropic form of the spherical Einstein-connected reference solution, that representative is mapped exactly to Euclidean space. Hence
\begin{equation}
 \widetilde R^{(N)}=0 .
 \label{eq:LovelockRNzero}
\end{equation}

The time-symmetric Lovelock Hamiltonian constraint involves
\begin{equation}
 {\cal Q}_L[h]
 =
 R[h]
 +\sum_{p=2}^{K}
 \alpha_p{\cal L}_p[h].
 \label{eq:LovelockConstraint}
\end{equation}
Equation~(\ref{eq:LovelockInductionAssumption}) implies $\widetilde{\rm Riem}=O(\epsilon^N)$. Since ${\cal L}_p$ contains $p$ powers of the curvature,
\begin{equation}
 \alpha_p\widetilde{\cal L}_p
 =
 O(\epsilon^{pN+1}),
 \qquad p\ge2 .
 \label{eq:LovelockOrderCounting}
\end{equation}
No higher Lovelock density therefore contributes at order $\epsilon^N$. Since all lower coefficients vanish by the induction hypothesis,
\begin{equation}
 \widetilde{\cal Q}_L
 =
 O(\epsilon^{N+1}) .
 \label{eq:LovelockQNzero}
\end{equation}

Let
\begin{equation}
 \widetilde h_{ij}
 =
 \delta_{ij}
 +\epsilon^N k^{(N)}_{ij}
 +O(\epsilon^{N+1})
 \label{eq:LovelockFirstUnfixed}
\end{equation}
at the first potentially non-vanishing order. Equation~(\ref{eq:LovelockZeroMass}) gives $\widetilde M_{\rm ADM}=0$ exactly, while Eq.~(\ref{eq:LovelockQNzero}) means that in the theory consistently truncated at $O(\epsilon^N)$ one has
\begin{equation}
 \widetilde{\cal Q}_L=0,
 \qquad
 \widetilde M_{\rm ADM}=0.
\end{equation}
The regular Taylor expansion is again crucial: a non-gauge $k^{(N)}_{ij}$ would be a genuine non-flat zero-mass equality configuration of the truncated theory, and terms of order $\epsilon^{N+1}$ and higher cannot remove that lower-order coefficient. It is therefore excluded by the assumed equality-case rigidity. Hence
\begin{equation}
 k^{(N)}_{ij}
 =
 {\cal L}_{\xi_N}\delta_{ij},
\end{equation}
and after an order-$\epsilon^N$ gauge transformation,
\begin{equation}
 \widetilde h_{ij}
 =
 \delta_{ij}
 +O(\epsilon^{N+1}).
 \label{eq:LovelockInductionStep}
\end{equation}
Since $N$ is arbitrary,
\begin{equation}
 \widetilde h_{ij}=\delta_{ij}
 \label{eq:LovelockFlatAllOrders}
\end{equation}
to all orders in the regular Taylor expansion.

We finally give the geometric step that turns conformal flatness into spherical symmetry. Regularity of a nondegenerate static bifurcate horizon gives $K_{AB}=0$ on the physical spatial slice. Because the branch-adapted conformal factor depends only on $V$, the transformed horizon satisfies
\begin{equation}
 \widetilde K_{AB}
 \propto
 \widetilde q_{AB},
\end{equation}
and is therefore a compact totally umbilical hypersurface in Euclidean space. It is consequently a round $S^{D-2}$.

It remains to show that this roundness propagates through the exterior. At the first possible nonspherical perturbative order, say $\epsilon^N$, the difference from the spherical reference solution obeys the homogeneous linearized Einstein equations. The coefficient governing the nonspherical projections can therefore be obtained from the GR conformal equations evaluated on the Schwarzschild--Tangherlini background.

Let
\begin{equation}
 \widetilde h^{(0)}_{ij}
 =
 U_0^{4/(D-3)}h^{(0)}_{ij}
 =\delta_{ij},
 \qquad
 U_0=\frac{1+V_0}{2}.
 \label{eq:EndBunting}
\end{equation}
Thus
\begin{equation}
 h^{(0)}_{ij}=e^{2\sigma_0}\delta_{ij},
 \qquad
 \sigma_0=-\frac{2}{D-3}\ln U_0,
\end{equation}
with
\begin{equation}
 \sigma_0'(V_0)
 =
 -\frac{2}{(D-3)(1+V_0)} .
 \label{eq:sigmaprime}
\end{equation}

Write $H_{ij}=\partial_i\partial_jV$ in the flat conformal geometry and let $H^{(N)}_{ij}$ denote its first unfixed nonspherical coefficient. For a conformally flat metric $h_{ij}=e^{2\sigma}\delta_{ij}$,
\begin{equation}
\begin{aligned}
 R_{ij}[h]
 ={}&
 -(D-3)\partial_i\partial_j\sigma
 -\delta_{ij}\partial^2\sigma
 \\
 &+(D-3)
 \left(
 \partial_i\sigma\,\partial_j\sigma
 -\delta_{ij}(\partial\sigma)^2
 \right),
\end{aligned}
\label{eq:RicciConformalFlat}
\end{equation}
and
\begin{equation}
\begin{aligned}
 D_iD_jV=H_{ij}-\partial_i\sigma\,\partial_jV-\partial_j\sigma\,\partial_iV+\delta_{ij}\partial^k\sigma\,\partial_kV .
\end{aligned}
\label{eq:HessianConformal}
\end{equation}
On a level surface of the background lapse $V_0$, with $A,B$ tangent to the surface and $n$ normal to it, $\partial_A V_0=\partial_A\sigma_0=0$. The tangential traceless and mixed parts of the homogeneous linearized static Einstein equation therefore contain the same background coefficient. Equivalently, linearizing the corresponding GR relations about the spherical background gives
\begin{equation}
 \left[-V_0(D-3)\sigma_0'(V_0)-1\right]
 \bigl(H_{AB}^{(N)}\bigr)^{\rm TF}=0,
\end{equation}
and
\begin{equation}
 \left[-V_0(D-3)\sigma_0'(V_0)-1\right]
 H_{nA}^{(N)}=0.
\end{equation}
The variation of the coefficient itself does not contribute because the background already satisfies $(H_{AB}^{(0)})^{\rm TF}=H_{nA}^{(0)}=0$. Using
Eq.~(\ref{eq:sigmaprime}),
\begin{equation}
 -V_0(D-3)\sigma_0'(V_0)-1
 =
 \frac{2V_0}{1+V_0}-1
 =
 \frac{V_0-1}{V_0+1}.
\end{equation}
Hence
\begin{equation}
 \frac{V_0-1}{V_0+1}
 \bigl(H_{AB}^{(N)}\bigr)^{\rm TF}=0,
 \qquad
 \frac{V_0-1}{V_0+1}H_{nA}^{(N)}=0.
 \label{eq:LovelockHessianConditions}
\end{equation}
Thus the relevant coefficient is
\begin{equation}
 \boxed{
 C(V_0)=\frac{V_0-1}{V_0+1}
 } .
 \label{eq:CVexplicit}
\end{equation}
For the black hole exterior, $0\leq V_0<1$ at every finite point, so $C(V_0)\neq0$ there; it vanishes only asymptotically as $V_0\to1$. Consequently,
\begin{equation}
 \bigl(H_{AB}^{(N)}\bigr)^{\rm TF}=0,
 \qquad
 H_{nA}^{(N)}=0 .
\end{equation}
The first relation removes the traceless deformation of the round level surfaces at the first nonspherical order, while the second removes the angular variation of their normal displacement. Beginning with the round horizon, the first possible nonspherical coefficient therefore vanishes. Repeating the same argument order by order leaves only concentric round level surfaces, and hence $V=V(r)$ to all orders in the Taylor expansion.

The spacetime is consequently spherically symmetric. For generic Lovelock couplings, the asymptotically flat Einstein-connected spherically symmetric vacuum solution is the branch determined by Eq.~(\ref{eq:LovelockPolynomial}) \cite{Wheeler1986,Zegers2005}. Thus, conditional on the corresponding Lovelock zero-mass positive-mass rigidity statement and away from degenerate choices of the Lovelock couplings, static black hole uniqueness extends perturbatively to all orders on the regular Taylor-expandable Einstein-connected Lovelock branch.

\end{document}